\documentclass[12pt,a4paper]{article}
\usepackage{amsmath}
\usepackage{amssymb}
\usepackage{cite}

\renewcommand{\thesection}
  {\arabic{section}.\hspace{-.5em}}

\makeatletter
\renewcommand\section{
  \@startsection{section}{3}{\z@}%
  {-3.25ex\@plus -1ex \@minus -.2ex}%
  {1.5ex \@plus .2ex}%
  {\normalfont\normalsize\bfseries\mathversion{bold}}}
\renewcommand\subsection{
  \@startsection{subsection}{3}{\z@}%
  {-3.25ex\@plus -1ex \@minus -.2ex}%
  {1.5ex \@plus .2ex}%
  {\normalfont\normalsize\bfseries\mathversion{bold}}}
\makeatother

\makeatletter \@addtoreset{equation}{section} \makeatother
\renewcommand{\theequation}{\arabic{section}.\arabic{equation}}

\makeatletter
\renewcommand{\appendix}{
\renewcommand{\thesection}{\Alph{section}.\hspace{-.5em}}
\@addtoreset{equation}{section}
\renewcommand{\theequation}{\Alph{section}.\arabic{equation}}
\setcounter{section}{0}}
\makeatother

\newcommand{\Eqn}[1]{&\hspace{-0.5em}#1\hspace{-0.5em}&}
\newcommand{\nn}{\nonumber}
\renewcommand{\[}{\begin{equation}}
\renewcommand{\]}{\end{equation}}
\newcommand{\eqb}{\begin{eqnarray}}
\newcommand{\eqe}{\end{eqnarray}}

\newcommand{\grp}[1]{\mathrm{#1}}
\newcommand{\bvec}[1]{\boldsymbol{#1}}
\newcommand{\varth}{\vartheta}

\newcommand{\bbR}{{\mathbb R}}
\newcommand{\bbZ}{{\mathbb Z}}

\newcommand{\mut}{\mu^\vee}
\newcommand{\bfR}{\boldsymbol{R}}

\newcommand{\scr}[1]{{\mbox{\scriptsize #1}}}
\newcommand{\sscr}[1]{{\mbox{\tiny #1}}}

\newcommand{\gen}{\scr{gen}}
\newcommand{\pert}{\scr{pert}}
\newcommand{\inst}{\scr{inst}}
\newcommand{\qinst}{\scr{$q$-inst}}
\newcommand{\YM}{\scr{YM}}
\newcommand{\UV}{\scr{UV}}
\newcommand{\denom}{\scr{denom}}

\newcommand{\Nf}{N_\scr{f}}
\newcommand{\phithere}{{\phi_\scr{there}}}

\begin{document}

%%%%%%%%%%%%%%%%%%%%%%%%%%%%%%%%%%%%%%%%%%%%%%%%%%%%%%%%%%%%%%%%%%%%%%%%
%%% Title page %%%

%%% styles for the title page %%%%%%%%%%%%%%%%%%%%%%%%%%%%%%%%%%%%%%%%%%
\def\papertitlepage{\baselineskip 3.5ex \thispagestyle{empty}}
\def\preprinumber#1#2{\hfill
\begin{minipage}{1.2in}
#1 \par\noindent #2
\end{minipage}}

%%%%%%%%%%%%%%%%%%%%%%%%%%%%%%%%%%%%%%%%%%%%%%%%%%%%%%%%%%%%%%%%%%%%%%%%
%
\papertitlepage
\setcounter{page}{0}
\preprinumber{arXiv:1408.3619}{}
\vskip 2ex
\vfill
\begin{center}
{\large\bf\mathversion{bold}
A reduced BPS index of E-strings
}
\end{center}
\vfill
\baselineskip=3.5ex
\begin{center}
Kazuhiro Sakai\\

{\small
\vskip 6ex
{\it Department of Physical Sciences, Ritsumeikan University}\\[1ex]
{\it Shiga 525-8577, Japan}\\
\vskip 1ex
{\tt ksakai@fc.ritsumei.ac.jp}

}
\end{center}
\vfill
\baselineskip=3.5ex
\begin{center} {\bf Abstract} \end{center}

We study the BPS spectrum of E-strings in a situation where the global
$E_8$ symmetry is broken down to $D_4\oplus D_4$ by a certain twist.
We find that the refined BPS index in this setup serves as a reduced BPS
index of E-strings, which gives a novel trigonometric generalization of
the Nekrasov partition function for four-dimensional ${\cal N}=2$
supersymmetric $\grp{SU}(2)$ gauge theory with $\Nf=4$ massless flavors.
We determine the perturbative part of this index and also first few
unrefined instanton corrections. By using these results together with
the modular anomaly equation, the genus expansion of the free energy can
be computed efficiently up to very high order. We also determine the
unrefined elliptic genus of four E-strings under the twist.

\vfill
\noindent
August 2014

%%%%%%%%%%%%%%%%%%%%%%%%%%%%%%%%%%%%%%%%%%%%%%%%%%%%%%%%%%%%%%%%%%%%%%%%
%%% Main text %%%

%%% styles for the main text %%%%%%%%%%%%%%%%%%%%%%%%%%%%%%%%%%%%%%%%%%%
\setcounter{page}{0}
\newpage
\renewcommand{\thefootnote}{\arabic{footnote}}
\setcounter{footnote}{0}
\setcounter{section}{0}
\baselineskip = 3.5ex
\pagestyle{plain}
%

%%%%%%%%%%%%%%%%%%%%%%%%%%%%%%%%%%%%%%%%%%%%%%%%%%%%%%%%%%%%%%%%%%%%%%%%
\section{Introduction}
%%%%%%%%%%%%%%%%%%%%%%%%%%%%%%%%%%%%%%%%%%%%%%%%%%%%%%%%%%%%%%%%%%%%%%%%

The E-string theory is one of the simplest
interacting supersymmetric field theories in six dimensions
\cite{Ganor:1996mu,Seiberg:1996vs,
      Klemm:1996hh,Ganor:1996pc,Minahan:1998vr}.
The theory arises from an M5-brane probing the end-of-the-world 9-plane
in the heterotic M-theory.
The fundamental objects of the theory are
called E-strings and are realized by
M2-branes stretched between the M5-brane and the 9-plane.
While the theory itself has begun to attract revived interests
\cite{Heckman:2013pva, Ohmori:2014pca},
the BPS spectrum of E-strings has been drawing continuous attention.
One reason for this is that the BPS spectrum of
toroidally compactified E-string theory
encompasses that of various gauge theories
with rank-one gauge groups in five and four dimensions
\cite{Ganor:1996pc}.
Another reason is that the BPS index of E-strings is 
essentially equivalent to the topological string partition function
for the local $\frac{1}{2}$K3 Calabi--Yau threefold
\cite{Klemm:1996hh, Minahan:1998vr},
which is a key example to study
topological strings on non-toric Calabi--Yau threefolds.

In this paper we focus on a particular twist of the E-string theory.
We consider a situation where
the global $E_8$ symmetry is broken down to $D_4\oplus D_4$.
In other words, we consider a Seiberg--Witten system
\cite{Seiberg:1994rs,Seiberg:1994aj}
which has one-dimensional Coulomb branch moduli space
with two $D_4$-type singularities.
This is very reminiscent of the moduli space of
four-dimensional ${\cal N}=2$ supersymmetric 
$\grp{SU}(2)$ gauge theory with $\Nf=4$ massless
flavors \cite{Seiberg:1994aj}.
In fact, concerning the moduli space the only difference is that
in our setup two $D_4$ singularities are placed at a finite distance
from each other, whereas in the $\grp{SU}(2)\ \Nf=4$ theory
one of the $D_4$ singularities is at infinity.
A peculiar feature common to these Seiberg--Witten systems is that
the complex structure of the Seiberg--Witten curve,
which represents the gauge coupling,
is constant all over the moduli space.
This leads to substantial simplification
of the BPS spectra.
Nevertheless, the resulting spectra are not boring.
In fact, it is well known that
the Nekrasov partition function
for the $\grp{SU}(2)\ \Nf=4$ theory
\cite{Nekrasov:2002qd,Nekrasov:2003rj}
is nontrivial even in the massless limit
\cite{Grimm:2007tm}.
The main objective of this paper is to clarify
the counterpart of this function on the E-string theory side.

More specifically, we study
the refined BPS index of E-strings
evaluated under the $D_4\oplus D_4$ twist.
We find that this
serves as a reduced BPS index of E-strings.
A characteristic feature of the reduced index is that
it inherits the modular properties from the original BPS index
of E-strings. In particular, the reduced index satisfies
the same modular anomaly equation
as that of the original BPS index
\cite{Hosono:1999qc,Huang:2013yta}.
There is another function which
also possesses such modular characteristics:
the refined BPS index of E-strings
with untwisted $E_8$ symmetry.
Interestingly, our reduced BPS index is even simpler than
this function and thus easier to deal with.
Hence the reduced index would serve as
a convenient tool
for understanding the whole structure
of the BPS spectrum of E-strings.

We study in detail this reduced BPS index
by performing three kinds of expansions.
One is the genus expansion of topological strings,
another is the $q$-expansion, which is closely related to
the spacetime instanton expansion, and the other
is the winding number expansion,
which generates elliptic genera of multiple E-strings.
Making use of the modular anomaly equation
and comparing these expansions with one another,
we are able to clarify the structure of the expansion
as well as to determine some expansion coefficients.
Among others, an intriguing outcome of this study is that
the reduced BPS index gives a novel trigonometric generalization of
the Nekrasov partition function for
4d ${\cal N}=2$ supersymmetric
$\grp{SU}(2)$ theory with $\Nf=4$ massless flavors.

The paper is organized as follows.
In section~2, we present the definition of the reduced BPS index
and summarize its general properties.
In section~3, we consider the genus expansion of the free energy
and study topological string amplitudes
for the local $\frac{1}{2}$K3
Calabi--Yau threefold
evaluated under the $D_4\oplus D_4$ twist.
We clarify the general structure of
these amplitudes and discuss how to determine them.
In section~4, we consider the instanton expansion of the free energy.
Our conjectures about the perturbative part
and first few unrefined instanton corrections are presented.
We discuss their structures, in particular
in relation to Nekrasov partition functions
for ordinary gauge theories.
In section~5, we consider the winding number expansion
and study elliptic genera of multiple E-strings.
The elliptic genus of two E-strings is presented in a very simple
form. We also discuss the structure of general unrefined elliptic
genera and determine the unrefined elliptic genus of four E-strings.
In section~6, we elucidate how the reduced BPS index reproduces
the Nekrasov partition function for 4d ${\cal N}=2\ \grp{SU}(2)$
gauge theory with $\Nf=4$ massless flavors.
In section~7 we discuss possible extensions of this work.
In Appendix~A, we summarize our conventions of special functions
and present some useful formulas.

%%%%%%%%%%%%%%%%%%%%%%%%%%%%%%%%%%%%%%%%%%%%%%%%%%%%%%%%%%%%%%%%%%%%%%%%
\section{Definition and general properties}
%%%%%%%%%%%%%%%%%%%%%%%%%%%%%%%%%%%%%%%%%%%%%%%%%%%%%%%%%%%%%%%%%%%%%%%%

%%%
\subsection{Brief review of refined BPS index of E-strings}
%%%

In this subsection we briefly review the refined BPS index of
E-strings.
Consider the E-string theory on $\bbR^5\times S^1$.
One can regard this theory
effectively as a 5d ${\cal N}=1$ supersymmetric field theory,
where BPS E-strings wound around $S^1$ are viewed as
5d BPS particles.
The refined BPS index of E-strings is defined as the 5d BPS index
\cite{Nekrasov:2002qd,Choi:2012jz}
for this toroidally compactified E-string theory:
\[\label{Zgendef}
Z^\gen
(\phi,\tau,\bvec{m},\epsilon_1,\epsilon_2)
:={\rm Tr}\, (-1)^{2J_\sscr{L}+2J_\sscr{R}}
 y_\scr{L}^{J_\sscr{L}}
 y_\scr{R}^{J_\sscr{R}+J_\sscr{I}}
 p^n q^k e^{i\bvec{\Lambda}\cdot\bvec{m}},
\]
where
\[
y_\scr{L}:=e^{i(\epsilon_1-\epsilon_2)},\qquad
y_\scr{R}:=e^{i(\epsilon_1+\epsilon_2)},\qquad
p:=e^{-\phi},\qquad q:=e^{2\pi i\tau}.
\]

Here $J_\scr{L},J_\scr{R},J_\scr{I}$
and $\bvec{\Lambda}=(\Lambda_1,\dots,\Lambda_8)$
are spins (or weights of the associated Lie algebras) of
the little group
$\grp{SO}(4)=\grp{SU}(2)_\scr{L}\times\grp{SU}(2)_\scr{R}$,
the R-symmetry group $\grp{SU}(2)_\scr{I}$
and the global symmetry group $E_8$ respectively.
Nonnegative 
integers $n,k$ are respectively
the winding number and the momentum along $S^1$.
As wee see the above index is a function in twelve variables:
$\phi$ is the tension of the E-strings,
$\tau$ is proportional to the inverse of the radius of $S^1$,
$\bvec{m}=(m_1,\ldots,m_8)$
and $\epsilon_1,\epsilon_2$ are respectively
the Wilson line parameters
for the global symmetries $E_8$ and $\grp{SO(4)}$.

The BPS index $Z^\gen$ can be viewed as the generating function
for the sequence of elliptic genera of multiple E-strings.
It is expanded as
\eqb
Z^\gen
 \Eqn{=}1+\sum_{n=1}^\infty p^nZ^\gen_n,
\eqe
where $Z^\gen_n$ is (the holomorphic limit of)
the elliptic genus of $n$ E-strings.
The elliptic genus of single E-strings is known as
\cite{Huang:2013yta}
\[
Z^\gen_1
=-q^{1/2}\frac{A_1(\tau,\bvec{m})}
 {\eta^6\varth_1(\epsilon_1)\varth_1(\epsilon_2)}.
\]
Here
$A_1(\tau,\bvec{m})
 :=\frac{1}{2}\sum_{k=1}^4\prod_{j=1}^8\varth_k(m_j)$
is the theta function of the $E_8$ root lattice.
The elliptic genus of two E-strings
$Z^\gen_2$
was also calculated recently \cite{Haghighat:2014pva}.
In general, 
$Z^\gen_n$ is some Weyl$(E_8)$-invariant
Jacobi form of index $n$
with respect to $\bvec{m}$
and can be expressed in terms of nine generators
\cite{Sakai:2011xg}:
$A_k(\tau,\bvec{m})$ with $k=1,2,3,4,5$
and $B_k(\tau,\bvec{m})$ with $k=2,3,4,6$,
where $k$ represents the index of the Jacobi form,
or, the level of the associated $E_8$ current algebra.

The BPS index can also be viewed as
the instanton part
of the refined topological string
partition function for
the local $\frac{1}{2}$K3 Calabi--Yau threefold:
\[
Z^\gen=\exp F^\gen
\]
with
\[
F^\gen=\sum_{n=0}^\infty\sum_{g=0}^\infty
 (\epsilon_1+\epsilon_2)^{2n}(-\epsilon_1\epsilon_2)^{g-1}
 F^{\gen(n,g)}.
\]
Topological string amplitudes
for the local $\frac{1}{2}$K3
at low genus have been studied extensively.
See e.g.~\cite{Sakai:2011xg,Huang:2013yta}
for the latest results of $F^{\gen(n,g)}$
with general $\bvec{m}$.
In general, a topological string amplitude at low genus
has its classical part,
which is
some polynomial at most cubic in the K\"ahler moduli parameters.
In this paper
we adopt (\ref{Zgendef})
as the primary definition of $Z^\gen$
and let $F^{\gen(n,g)}$ be made up of
purely world-sheet instanton contributions.

%%%
\subsection{Reduced BPS index and $D_4\oplus D_4$ twist}
%%%

In this paper we consider a BPS index of E-strings defined as
\[
Z(\phi,\tau,\epsilon_1,\epsilon_2)
:={\rm Tr}\,
 (-1)^{2J_\sscr{L}+2J_\sscr{R}
      +\Lambda_3+\Lambda_4-\Lambda_5-\Lambda_6}
 y_\scr{L}^{J_\sscr{L}}y_\scr{R}^{J_\sscr{R}+J_\sscr{I}}
 p^n q^{k+(-\Lambda_5-\Lambda_6+\Lambda_7+\Lambda_8)/2}.
\]
This is equivalent to the refined BPS index (\ref{Zgendef}) evaluated
at the following special values of the $E_8$ Wilson line parameters
\[\label{muD4D4}
m_1=m_2=0,\qquad
m_3=m_4=\pi,\qquad
m_5=m_6=-\pi-\pi\tau,\qquad
m_7=m_8=\pi\tau.
\]
This setting of Wilson line parameters corresponds
to breaking the global symmetry $E_8$ down to
$D_4\oplus D_4$.
This is manifestly seen in the Seiberg--Witten description:
The Seiberg--Witten curve for the E-string theory
\cite{Eguchi:2002fc}
with the above parameter values reduces
to the following very simple form \cite{Eguchi:2002nx}
\eqb
\label{WScurve}
y^2\Eqn{=}
  4x^3-\frac{E_4}{12}\left(u^2-u_0^2\right)^2 x
      -\frac{E_6}{216}\left(u^2-u_0^2\right)^3,
\eqe
where
\[
u_0:=-\frac{2}{q^{1/2}\eta^{12}}.
\]
The curve describes an elliptic fibration over the moduli space
parametrized by $u$,
with $D_4$ type singularities at $u=\pm u_0$.
It is easy to see that the complex structure of the above curve is
$\tau$ and does not depend on $u$.

As we mentioned, the refined BPS index of E-strings
is expressed in terms of nine Weyl$(E_8)$-invariant Jacobi forms
$A_k,B_k$.
Under the twist (\ref{muD4D4}), they become
\begin{align}
\hspace{2em}
A_1&=A_3=A_5=B_3=0,\hspace{-10em}&&&&\nn\\[1ex]
A_2&= \frac{E_4}{9q},&
A_4&= \frac{E_4}{q^2},&&&\nn\\
B_2&=-\frac{E_6}{15q},&
B_4&=-\frac{E_6}{15q^2},&
B_6&= \frac{E_6}{9q^3}.
\label{ABforms}
\hspace{2em}
\end{align}
There are several known results about $Z^\gen$
expressed in terms of these Jacobi forms
\cite{Sakai:2011xg,Huang:2013yta,Haghighat:2014pva}.
These results can be immediately translated
for our reduced BPS index $Z$
by simply substituting (\ref{ABforms}).
We will do this in the following sections.
Note that all Jacobi forms
of odd index vanish under the twist.
Nonvanishing Jacobi forms also take very simple forms.
This leads to substantial simplification
of the refined BPS index.

%%%
\subsection{Expansions}
%%%

In the following sections we will study the index $Z$
by performing three kinds of expansions.
First, as in the case of $Z^\gen$, the free energy
\[
F=\ln Z
\]
admits the genus expansion
\[\label{genusexp}
F=\sum_{n=0}^\infty\sum_{g=0}^\infty
 (\epsilon_1+\epsilon_2)^{2n}(-\epsilon_1\epsilon_2)^{g-1}
 F^{(n,g)}(\phi,\tau).
\]
Here
$F^{(n,g)}$ are interpreted as
topological string amplitudes evaluated
under the twist (\ref{muD4D4}).
We will study them in section~3.

It is also interesting to consider
the instanton expansion defined as follows
\[\label{instexp}
F=F^\pert(\phi,\epsilon_1,\epsilon_2)
  +\sum_{k=1}^\infty q^k F^\qinst_k(\phi,\epsilon_1,\epsilon_2).
\]
This expansion is closely related to the
instanton expansion of conformal gauge theories.
We will study this expansion in section~4.

The other expansion is the winding number expansion
\eqb\label{ellgenexp}
Z\Eqn{=}1+\sum_{n=1}^\infty p^{2n}Z_{2n}(\tau,\epsilon_1,\epsilon_2).
\eqe
$Z_{2n}$ is, as before, the elliptic genus of $2n$ E-strings
evaluated under the $D_4\oplus D_4$ twist.
$Z_k$ with odd $k$ are absent because
Weyl$(E_8)$-invariant Jacobi forms of odd index vanish
under the $D_4\oplus D_4$ twist,
as we saw in (\ref{ABforms}).
We will study these elliptic genera in section~5.

%%%
\subsection{Modular anomaly equation}
%%%

The BPS index $Z^\gen$ for E-strings is known to
satisfy the modular anomaly equation
\cite{Hosono:1999qc,Huang:2013yta}.
As one can see from (\ref{ABforms}),
the $D_4\oplus D_4$ twist does not modify
the structure of modular anomalies,
i.e.~the way how $E_2$ enters in the index.
Thus, our reduced BPS index $Z$
satisfies the same modular anomaly equation
\[
\partial_{E_2}Z
=-\frac{1}{24}\left[
 \epsilon_1\epsilon_2\partial_\phi\left(\partial_\phi-1\right)
+(\epsilon_1+\epsilon_2)^2\partial_\phi
\right]Z.
\]
In terms of $F^{(n,g)}$, it is expressed as
\eqb\label{MAEforFng}
\partial_{E_2}F^{(n,g)}
\Eqn{=}\frac{1}{24}\sum_{n_1=0}^n\sum_{g_1=0}^g
  \partial_\phi F^{(n_1,g_1)}\partial_\phi F^{(n-n_1,g-g_1)}\nn\\
&&+\frac{1}{24}\partial_\phi\left(\partial_\phi-1\right)F^{(n,g-1)}
  -\frac{1}{24}\partial_\phi F^{(n-1,g)},
\eqe
where $F^{(n,g)}=0$ if $n<0$ or $g<0$.
We will use this form of the equation in section~3.
In terms of the elliptic genera $Z_{2n}$
the modular anomaly equation is simply
\[\label{MAEforZn}
\partial_{E_2}Z_{2n}=\frac{1}{24}
\left[-2n(2n+1)\epsilon_1\epsilon_2
 +2n(\epsilon_1+\epsilon_2)^2\right]Z_{2n},
\]
which we will use in section~5.

%%%%%%%%%%%%%%%%%%%%%%%%%%%%%%%%%%%%%%%%%%%%%%%%%%%%%%%%%%%%%%%%%%%%%%%%
\section{Genus expansion}
%%%%%%%%%%%%%%%%%%%%%%%%%%%%%%%%%%%%%%%%%%%%%%%%%%%%%%%%%%%%%%%%%%%%%%%%

In this section we study the genus expansion (\ref{genusexp}).
The expansion coefficients $F^{(n,g)}$ represent
topological string amplitudes for the local $\frac{1}{2}$K3
evaluated under the $D_4\oplus D_4$ twist.
Explicit forms of general amplitudes
$F^{\gen(n,g)}$ at low genus are known
\cite{Sakai:2011xg}.
There $F^{\gen(n,g)}$ are expressed in terms of
period integrals over the Seiberg--Witten curve,
which are related to
the coordinate of the moduli space by the mirror map.
Thus, we first summarize in subsection~3.1
how the mirror map
and the period integrals are reduced
under the $D_4\oplus D_4$ twist.

To determine higher genus amplitudes,
the modular anomaly equation (\ref{MAEforFng})
works as a powerful tool.
In subsection~3.2, 
we derive the general structure of $F^{(n,g)}$
and explain how to determine the modular ambiguities
of $F^{(n,g)}$ at high genus.

%%%
\subsection{Mirror map}
%%%

The purpose of this subsection is to present 
how the mirror map and the period integrals studied in
\cite{Sakai:2011xg} are reduced
under the $D_4\oplus D_4$ twist.
We intend this subsection
to be a short summary of the results.
The reader is referred for definitions of the symbols
used here and in (\ref{FhK310}) to
\cite{Sakai:2011xg}. Note also that the variable
$\phi$ used in \cite{Sakai:2011xg} should not be confused
with $\phi$ in this paper. They are related with each other by
\[
\phithere
=-\phi_{\mbox{\scriptsize here}}
 +\ln\bigl(-q^{1/2}\eta^{12}\bigr).
\]

As we mentioned, the gauge coupling $\tilde\tau(u)$
of the low energy effective theory
is constant all over the moduli space
\[
\tilde\tau=\tau.
\]
Due to this, the mirror map takes a very simple form.
One obtains
\eqb
\omega\Eqn{=}\left(u^2-u_0^2\right)^{-1/2},\\
p\Eqn{=}e^{-\phi}
=\frac{u_0}{2}e^\phithere
=\frac{u_0}{u+\sqrt{u^2-u_0^2}}.
\eqe
These relations can also be written as
\eqb
u\Eqn{=}u_0\cosh\phi,\\
\omega\Eqn{=}\frac{1}{u_0\sinh\phi}.
\eqe
It then follows that
\eqb
\label{ttautau}
\tilde\Delta\Eqn{=}\Delta,\qquad \tilde{E}_{2n}=E_{2n},\\
\label{omegainphi}
\ln\omega\Eqn{=}\phithere-\ln(1-p^2).
\eqe
%

%%%
\subsection{Topological string amplitudes}
%%%

In \cite{Sakai:2011xg},
explicit forms of $F^{\gen(n,g)}$
with $n=0,\ 0\le g\le 3$ were calculated.
It is straightforward to generalize the calculation
to the refined case.
For our present purposes
it is sufficient to compute the explicit form of $F^{\gen(1,0)}$
in addition.
The result is
\[\label{FhK310}
 F^{\gen(1,0)}
=\frac{1}{2}\ln\omega-\frac{1}{24}\ln\tilde\Delta
 +\frac{1}{24}\ln\Delta-\frac{1}{2}\phithere.
\]
Using the relations (\ref{ttautau}), (\ref{omegainphi}),
one can translate these results of $F^{\gen(n,g)}$ into
\eqb
F^{(0,0)}\Eqn{=}0,\nn\\
F^{(0,1)}\Eqn{=}F^{(1,0)}=-\frac{1}{2}\ln\left(1-e^{-2\phi}\right),\nn\\
F^{(0,2)}\Eqn{=}\frac{1}{32}\frac{E_2}{\sinh^2\phi},\nn\\
F^{(0,3)}\Eqn{=}\frac{1}{768}\frac{3E_2^2+E_4}{\sinh^2\phi}
 +\frac{1}{384}\frac{2E_2^2+E_4}{\sinh^4\phi}.
\label{F0g}
\eqe
These results give us sufficient information to derive
the general structure of $F^{(n,g)}$.
Given the modular anomaly equation (\ref{MAEforFng}) and
the explicit forms of $F^{(0,0)}$, $F^{(0,1)}$, $F^{(1,0)}$,
it is not difficult to show that
$F^{(n,g)}$ with $n+g\ge 2$ has to take the following form
\eqb\label{Fngform}
F^{(n,g)}(\phi,\tau)\Eqn{=}\sum_{h=1}^{n+g-1}
 \frac{F^{(n,g,h)}(\tau)}{\sinh^{2h}\phi},
\eqe
where
\eqb
F^{(n,g,h)}(\tau)
\Eqn{=}\mbox{[quasi modular form of weight $2n+2g-2$]}.
\eqe
As an illustration, we present the explicit forms of 
$F^{(n,g)}$ with $n+g=2,3$:
\eqb
\label{Fngupto2}
F^{(2,0)}\Eqn{=}\frac{1}{96}\frac{E_2}{\sinh^2\phi},\qquad
F^{(1,1)}=\frac{1}{24}\frac{E_2}{\sinh^2\phi},\qquad
F^{(0,2)}=\frac{1}{32}\frac{E_2}{\sinh^2\phi},\\
\label{Fngupto3}
F^{(3,0)}\Eqn{=}\frac{1}{11520}\frac{5E_2^2+7E_4}{\sinh^2\phi}
 +\frac{1}{11520}\frac{5E_2^2+13E_4}{\sinh^4\phi},\nn\\
F^{(2,1)}\Eqn{=}\frac{1}{11520}\frac{35E_2^2+37E_4}{\sinh^2\phi}
 +\frac{1}{2880}\frac{10E_2^2+17E_4}{\sinh^4\phi},\nn\\
F^{(1,2)}\Eqn{=}\frac{1}{3840}\frac{25E_2^2+17E_4}{\sinh^2\phi}
 +\frac{1}{11520}\frac{95E_2^2+94E_4}{\sinh^4\phi},\nn\\
F^{(0,3)}\Eqn{=}\frac{1}{768}\frac{3E_2^2+E_4}{\sinh^2\phi}
 +\frac{1}{384}\frac{2E_2^2+E_4}{\sinh^4\phi}.
\eqe
We checked that these results are consistent with
the previous results \cite{Huang:2013yta}
about refined topological string amplitudes
for the local $\frac{1}{2}$K3 at low genus.

Let us next study how much
the modular anomaly equation (\ref{MAEforFng})
constrains the forms of $F^{(n,g)}$.
First, $F^{(n,g)}$ with $n+g=2$, presented in (\ref{Fngupto2}),
are uniquely determined by the modular anomaly equation,
given the explicit forms of $F^{(n,g)}$ with $n+g <2$.
$F^{(n,g)}$ with $n+g\ge 3$ are not determined solely by
the modular anomaly equation.
As we mentioned, $F^{(n,g,h)}(\tau)$ is
a quasi modular form of weight $2n+2g-2$.
Its anomalous part is completely determined
by the modular anomaly equation,
but its modular part contains
$\lceil(n+g)/6\rceil$
(or $\lceil(n+g)/6\rceil-1$ if $n+g\equiv 2$ mod 6)
undetermined coefficients.

There are several ways to fix these coefficients.
For example, in the next section we will study
$q$-expansion of the BPS index.
Full information on 
the leading order part $F^\pert$ completely
fixes the forms
of $F^{(n,g,h)}$ with $n+g\le 6$,
each of which contains at most
one undetermined coefficient.
We expect that the data of 
higher order coefficients $F^\qinst_k$
with $k\le N$, together with $F^\pert$,
will fix $F^{(n,g,h)}$ with $n+g\le 6(N+1)$.
We check it for $N=2$ in the unrefined case,
namely, we are able to determine $F^{(0,g,h)}$ with $g\le 18$.
Using other data one can fix different slices of
the BPS index.
We take the results of the following sections
in advance and summarize
how far we are able to determine $F^{(n,g,h)}$ at present:
\begin{itemize}
\item 
$F^{(n,g,h)}$ with $n+g\le 6$ and any $h$ are determined
using the explicit forms of $F^\pert$
presented in section~4.

\item 
$F^{(0,g,h)}$ with $g\le 18$ and any $h$ are determined
using the explicit forms of $F^\pert$, $F^\qinst_1$, $F^\qinst_2$
in the unrefined case presented in section~4.

\item
$F^{(n,g,1)}$ with any $n,g$
can be computed by expanding the elliptic genus
of two E-strings $Z_2$ presented in section~5.

\item
$F^{(0,g,2)}$ with any $g$ can be computed
from the elliptic genera $Z_2,Z_4$ in the unrefined case
presented in section~5.

\item
$F^{(n,g,n+g-1)}$ with any $n,g$ can be determined
by the Nekrasov partition function for the $\Nf=4$ theory
as we will see in section~6.

\end{itemize}
We do not present a vast amount of results here.
The computations are straightforward.
In doing this, formulas
(\ref{thetasigma})--(\ref{cnrecur}) are useful.

%%%%%%%%%%%%%%%%%%%%%%%%%%%%%%%%%%%%%%%%%%%%%%%%%%%%%%%%%%%%%%%%%%%%%%%%
\section{Instanton expansion}
%%%%%%%%%%%%%%%%%%%%%%%%%%%%%%%%%%%%%%%%%%%%%%%%%%%%%%%%%%%%%%%%%%%%%%%%

In this section we study the expansion (\ref{instexp}).
This expansion is closely related to the instanton expansion
of conformal gauge theories.
We first analyze the leading order part in subsection~4.1
and then study instanton corrections in subsection~4.2.

%%%
\subsection{Perturbative part}
%%%

$F^\pert$ is the leading order coefficient of the
$q$-expansion (\ref{instexp}).
We call it the perturbative part by analogy with that of
the Nekrasov partition function for conformal gauge theories.
It can also be viewed as the following limit
of the free energy
\[
F^\pert=\lim_{q\to 0}F.
\]
Since $\tau$ is proportional to the inverse of the radius
of $S^1$ around which E-strings are wound,
the above limit corresponds to the five-dimensional limit of
the E-string theory on $\bbR^5\times S^1$
under the $D_4\oplus D_4$ twist.

Let us first describe our result.
We make a conjecture that $F^\pert$ takes the following form
\[\label{q0Frefined}
F^\pert
 =-\sum_{n=1}^\infty
  \frac{\sin^2\left(n\epsilon_1/2\right)
       +\sin^2\left(n\epsilon_2/2\right)
       +\sin^2\left(n(\epsilon_1+\epsilon_2)/2\right)
       }
       {n\sin(n\epsilon_1)\sin(n\epsilon_2)}e^{-2n\phi}.
\]
Let us outline how we arrive at this form.
First, the leading order part
(i.e.~the summand with $n=1$) of (\ref{q0Frefined})
immediately follows from the explicit form of
$Z_2$, which we will see in the next section.
Next, by expanding the explicit forms (\ref{F0g}) of the
unrefined amplitudes $F^{(0,g)}$
in $p=e^{-\phi}$,
it is not difficult to notice that
$F^\pert$ in the unrefined case
must take the form
\eqb
\label{q0Funrefined}
F^\pert(\phi,\hbar,-\hbar)
\Eqn{=}\sum_{n=1}^\infty\frac{1}{2n\cos^2\left(n\hbar/2\right)}
  e^{-2n\phi}.
\eqe
Given the above two results, 
(\ref{q0Frefined}) arises as the most natural form of $F^\pert$.

We checked this conjecture against
the genus expansion of full $Z^\gen_4$ in the unrefined case
computed by the conventional method.
More importantly, the above form
reproduces the perturbative part of the Nekrasov partition function
for the $\grp{SU}(2)\ \Nf=4$ theory, as we will show in section~6.
The validity of the conjecture is therefore checked
in two extreme limits: $\phi\to\pm\infty$ and $\phi\to 0$.
This gives strong evidence to the conjecture.
Nevertheless, it would still be very interesting
if one could give a proof or a derivation for it.

It is useful to note that
the above $F^\pert$ can be expressed as
\eqb\label{Fpertingamma}
F^\pert
 \Eqn{=}
 -\gamma_{\epsilon_1,\epsilon_2}(2\phi)
 -\gamma_{\epsilon_1,\epsilon_2}(2\phi-i\epsilon_1-i\epsilon_2)
 +8\gamma_{2\epsilon_1,2\epsilon_2}(2\phi-i\epsilon_1-i\epsilon_2),
\eqe
where
\[\label{gamma5d}
\gamma_{\epsilon_1,\epsilon_2}(x)
 :=\sum_{n=1}^\infty\frac{1}{n}
  \frac{e^{-nx}}{\left(1-e^{in\epsilon_1}\right)
                 \left(1-e^{in\epsilon_2}\right)}.
\]
It is easy to see that $\gamma_{\epsilon_1,\epsilon_2}(x)$
can be rewritten as
\[
\gamma_{\epsilon_1,\epsilon_2}(x)
=\ln\prod_{k,l=1}^\infty
 \left(1-q_1^{k-1}t_2^l e^{-x}\right)
\]
with
\[
q_1:=e^{i\epsilon_1},\qquad t_2:=e^{-i\epsilon_2},
\]
when $|q_1|<1,\ |t_2|<1$.
By using this expression, the perturbative part can also be written as
\[\label{Zpert}
Z^\pert=\exp\left(F^\pert\right)
=\frac
 {\left[\prod_{k,l=1}^\infty
        \left(1-q_1^{2k-1}t_2^{2l-1} p^2\right)\right]^8}
 {\left[\prod_{k,l=1}^\infty\left(1-q_1^{k-1} t_2^l p^2\right)\right]
 \left[\prod_{k,l=1}^\infty\left(1-q_1^k t_2^{l-1} p^2\right)\right]}.
\]
One can clearly see that the perturbative part of the BPS index
is made up of the topological string partition functions
for the resolved conifold \cite{Iqbal:2007ii}.
The above expression is very reminiscent of
the perturbative part of the five-dimensional extension of 
the Nekrasov partition function for $\grp{SU}(2)$
theory with $\Nf=4$ massless flavors.
In fact, the denominator of the above $Z^\pert$
coincides with that of the perturbative part of the
5d Nekrasov partition function.
However, the numerator of the latter function is
$[\prod_{k,l=1}^\infty (1-q_1^{k-1/2}t_2^{l-1/2} p)]^8$,
which differs from the one we see in (\ref{Zpert}).\footnote{
Though the difference is small in appearance, it implies a qualitative
difference in the low energy effective gauge theory.
$F^\pert$ for the E-string theory under the $D_4\oplus D_4$ twist
yields a vanishing prepotential at genus zero,
meaning that the effective gauge
coupling does not vary. On the other hand,
the perturbative part of the Nekrasov partition function 
for 5d $\grp{SU}(2)\ \Nf=4$ theory
gives a nontrivial genus zero part,
meaning that the effective coupling depends on
the expectation value $\phi$.}
Interestingly, despite this difference,
both trigonometric generalizations reduce to
the same four-dimensional perturbative part,
as we will see in section~6.

%%%
\subsection{Instanton part}
%%%

Let us next consider instanton corrections.
For the sake of simplicity
in the rest of this section we restrict ourselves
to the unrefined case, namely, we focus on
\[
F^\qinst_n = 
F^\qinst_n(\phi,\hbar,-\hbar).
\]
As we mentioned in subsection~3.2, $F^\pert$ completely determines
the forms of $F^{(0,g)}$ with $g\le 6$.
By expanding them in $q$,
one obtains information about instanton corrections.
In fact, the above data
suffice to determine the form of the first
instanton correction $F^\qinst_1$.
This further gives us the data of $F^{(0,g)}$ with $g\le 12$.
Using these data (with several consistency conditions),
we are also able to determine $F^\qinst_2$.
Here we present the results:
\eqb
\label{Fqinst1}
F^\qinst_1
   \Eqn{=}\frac{t(4t-3)y+t^3}{(1-t)(y+t)^2},\\[1ex]
\label{Fqinst2}
F^\qinst_2
   \Eqn{=}\left[
 2t(6t-1)(2t-1)^2(4t-3)^2y^5\right.\nn\\
&& +t^2(2048t^7-10240t^6+23552t^5-29824t^4\nn\\
&&\hspace{2em}
      +21600t^3-8776t^2+1821t-144)y^4\nn\\
&& +t^3(8192t^7-40960t^6+90112t^5-109696t^4\nn\\
&&\hspace{2em}
      +78288t^3-32336t^2+7074t-621)y^3\nn\\
&& +t^4(28672t^7-135168t^6+272384t^5-302080t^4\nn\\
&&\hspace{2em}
      +197804t^3-75912t^2+15651t-1314)y^2\nn\\
&& +t^5(1024t^5-2944t^4
      +3288t^3-1736t^2+408t-27)(4t-3)^2y\nn\\
&&\left. +2t^7(6t^2-8t+3)(4t-3)^4
\right]\nn\\
&&
\big/\left[2(1-t)(2t-1)^2(y+t)^4(y+t(4t-3)^2)^2\right],
\eqe
where
\[
y:=\sinh^2 \phi,\qquad t:=\sin^2\frac{\hbar}{2}.
\]

As we mentioned repeatedly,
the BPS index $Z$
is regarded as the trigonometric generalization of
the Nekrasov partition function
for the 4d $\grp{SU}(2)\ \Nf=4$ theory.
In other words,
$Z$ reduces to the Nekrasov partition function
by taking an appropriate limit.
In this limit,
$F^\qinst_k$ reduce to their counterparts,
which will be denoted by $F^\qinst_{\YM,k}$.
By analyzing the Nekrasov partition function,
it can be easily verified that $F^\qinst_{\YM,k}$ determines
the contribution of up to and including $2k+1$ instantons.
We thus expect that our results
(\ref{q0Frefined}), (\ref{Fqinst1}) and (\ref{Fqinst2})
give the full information of
up to and including five `instanton' contributions
in the unrefined case.

Note that
the denominators of the above $F^\qinst_k$
can be expressed as
\eqb
\left[F^\qinst_1\right]_\denom
\Eqn{=}
 {\cos^2\frac{\hbar}{2}
  \left(\sinh^2\phi+\sin^2\frac{\hbar}{2}\right)^2},
\\[1ex]
\left[F^\qinst_2\right]_\denom
\Eqn{=}
 {2\cos^2\frac{\hbar}{2}\cos^2\frac{2\hbar}{2}
  \left(\sinh^2\phi+\sin^2\frac{\hbar}{2}\right)^4
  \left(\sinh^2\phi+\sin^2\frac{3\hbar}{2}\right)^2}.
\hspace{2em}
\eqe
This gives us a hint
about how to generalize the 4d Nekrasov partition function
to construct a closed Nekrasov-type expression for the BPS index $Z$.

%%%%%%%%%%%%%%%%%%%%%%%%%%%%%%%%%%%%%%%%%%%%%%%%%%%%%%%%%%%%%%%%%%%%%%%%
\section{Winding number expansion}
%%%%%%%%%%%%%%%%%%%%%%%%%%%%%%%%%%%%%%%%%%%%%%%%%%%%%%%%%%%%%%%%%%%%%%%%

In this section we consider the expansion (\ref{ellgenexp})
and study elliptic genera of multiple E-strings.
The modular anomaly equation (\ref{MAEforZn})
works as a powerful tool in determining the form of the elliptic genera.
It is expected that elliptic genera of multiple E-strings
are written in terms of Jacobi theta functions and elliptic functions
in $\epsilon_1,\epsilon_2$
\cite{Haghighat:2014pva}.
Recall that
$E_2$ never appears
in the expansions of Weierstrass elliptic functions.
The only source of the modular anomaly is the $E_2$
appearing in the theta function.
See (\ref{sigma_zexp})--(\ref{cnrecur}) and
(\ref{thetasigma}) for further details.

Let us first consider the leading coefficient $Z_2$.
The general elliptic genus of two E-strings
$Z^\gen_2$ was calculated recently \cite{Haghighat:2014pva}.
Thus the explicit form of $Z_2$ is derived
by simply substituting (\ref{ABforms}) into it.
We find that the result can be expressed
in a very simple form as follows:
\eqb
Z_2
 \Eqn{=} \frac{\varth_1(\epsilon_1)
               \varth_1(\epsilon_2)
               \varth_1(\epsilon_1+\epsilon_2)}
              {\eta^3
               \varth_1(2\epsilon_1)
               \varth_1(2\epsilon_2)}
         \frac{\wp'(\epsilon_1)-\wp'(\epsilon_2)}
              {\wp(\epsilon_1)-\wp(\epsilon_2)}\\[1ex]
 \Eqn{=}2\frac{\varth_1(\epsilon_1)
               \varth_1(\epsilon_2)
               \varth_1(\epsilon_3)}
              {\eta^3
               \varth_1(2\epsilon_1)
               \varth_1(2\epsilon_2)}
 \left(\zeta(\epsilon_1)+\zeta(\epsilon_2)+\zeta(\epsilon_3)\right),
\eqe
where
\[
\epsilon_3 := -\epsilon_1-\epsilon_2.
\]
Here $\zeta(z),\wp(z)$ are
Weierstrass elliptic functions (see Appendix~A).
The above explicit form completely determines $F^{(n,g,1)}$
(defined for $n+g\ge 2$) through
\[
Z_2=4\sum_{n,g=0}^\infty
(\epsilon_1+\epsilon_2)^{2n}(-\epsilon_1\epsilon_2)^{g-1}
F^{(n,g,1)}.
\]
In the unrefined case, the above expression
further reduces to a remarkably simple form:
\[
Z_2(\tau,\hbar,-\hbar)=2\frac{\varth_1(\hbar)^2}{\varth_1(2\hbar)^2}.
\]

Let us next consider elliptic genera $Z_{2n}$ with $n\ge 2$.
For the sake of simplicity in the rest of this subsection
we only consider the unrefined case
\[
Z_{2n}=Z_{2n}(\tau,\hbar,-\hbar).
\]
We make the following ansatz for the general form of $Z_{2n}$
in the unrefined case:
\eqb
Z_{2n}
 \Eqn{=}\frac{1}{\eta^{8(n^3-n)}}
 \frac{\varth_1(\hbar)^{(8n^3-2n)/3}}
      {\prod_{k=1}^n\varth_1(2k\hbar)^2}
        \Bigl[\mbox{polynomial in
        $\wp(\hbar),E_4,E_6$ of weight
        $\frac{8(n^3-n)}{3}$}\Bigr].\nn\\
\eqe
The denominator of the ansatz is inferred naturally from
the pole structure of $F^\pert$ given in the last section.
The rest of the ansatz is determined by
the modular anomaly equation (\ref{MAEforZn})
and other expected modular properties.

For $n=2$, the above ansatz contains 10 unknown coefficients.
By using the data of $F^\pert$ and $F^\qinst_1$
given in the last section, these coefficients are completely
fixed. We thus obtain
\eqb
Z_4
 \Eqn{=}\frac{1}{2\eta^{48}}
  \frac{\varth_1(\hbar)^{20}}
       {\varth_1(4\hbar)^2\varth_1(2\hbar)^2}
  \left(72\wp'^4\wp^2-18\wp''^2\wp'^2\wp+2\wp''\wp'^4+\wp''^4\right),
\eqe
where
$\wp=\wp(\hbar),\wp'=\wp''(\hbar),\wp''=\wp''(\hbar)$.
It turns out that the polynomial part is written in a concise
form by using $\wp,\wp'',\wp'^2$
as the generators instead of $\wp,E_4,E_6$.
The above result of $Z_4$ passed several nontrivial consistency checks.

For $n=3$,
the above ansatz for $Z_6$ contains 102 unknown coefficients.
The explicit forms of $F^\pert$ and $F^\qinst_k$ with $k=1,2$,
which are the best available data at present,
are not enough to determine $Z_6$ completely:
these data give 78 relations among the coefficients,
but 24 parameters are left to be determined by other means.

%%%%%%%%%%%%%%%%%%%%%%%%%%%%%%%%%%%%%%%%%%%%%%%%%%%%%%%%%%%%%%%%%%%%%%%%
\section{Relation to 4d ${\cal N}=2\ \grp{SU}(2)\ \Nf=4$ theory}
%%%%%%%%%%%%%%%%%%%%%%%%%%%%%%%%%%%%%%%%%%%%%%%%%%%%%%%%%%%%%%%%%%%%%%%%

The reduced BPS index $Z$ can be viewed as
a trigonometric generalization of the Nekrasov partition function
for 4d ${\cal N}=2\ \grp{SU}(2)$ gauge theory
with $\Nf=4$ massless flavors.
This means that $Z$ reduces to the Nekrasov partition function
in a certain limit.
In this section we elucidate how this occurs.
It was shown in \cite{Sakai:2012ik} that the Seiberg--Witten curve
for the E-string theory reproduces that for the $\Nf=4$ theory
by taking an appropriate limit.
Here we generalize this correspondence to the level of
the refined BPS index.

%%%
\subsection{Nekrasov partition function}
%%%

The Nekrasov partition function for
4d ${\cal N}=2$ supersymmetric $\grp{SU}(2)$ gauge theory
with $\Nf=4$ massless flavors is given
(up to a `$\grp{U}(1)$ factor' which we will mention later)
by
\cite{Nekrasov:2002qd,Nekrasov:2003rj}
\eqb
Z_\YM
\Eqn{=}Z_\YM^\pert Z_\YM^\inst.
\eqe
The perturbative part is given as
$Z_\YM^\pert=\exp F_\YM^\pert$ with
\eqb\label{FYMpert}
F_\YM^\pert
\Eqn{=}
 -\gamma^\YM_{\epsilon_1,\epsilon_2}(2a)
 -\gamma^\YM_{\epsilon_1,\epsilon_2}(2a-\epsilon_1-\epsilon_2)
 +8\gamma^\YM_{\epsilon_1,\epsilon_2}
   \left(a-\frac{\epsilon_1}{2}-\frac{\epsilon_2}{2}\right),
\eqe
where
\[
\gamma^\YM_{\epsilon_1,\epsilon_2}(x)
 :=\frac{d}{ds}\biggm|_{s=0}
  \frac{1}{\Gamma(s)}
  \int_0^\infty\frac{dt}{t}t^s
  \frac{e^{-tx}}{\left(e^{\epsilon_1 t}-1\right)
                 \left(e^{\epsilon_2 t}-1\right)}.
\]
The instanton part is given by
\eqb
\label{ZYMinst}
Z_\YM^\inst
\Eqn{=}\sum_{\bfR}
q_0^{|\bfR|}
\prod_{k,l=1}^2
\prod_{(i,j)\in R_k}
\prod_{\alpha=1}^2
\frac
{a_k+(j-\frac{1}{2})\epsilon_1
    +(i-\frac{1}{2})\epsilon_2}
{a_k-a_l+(\mu_{k,i} -j+\delta_{1,\alpha})\epsilon_1
        -(\mut_{l,j}-i+\delta_{2,\alpha})\epsilon_2},\qquad
\eqe
where
\[
a_1=a,\qquad a_2=-a.
\]
Here $\bfR=(R_1,R_2)$ denotes a pair of partitions
and $|\bfR|$
the total number of boxes in the Young diagrams of $R_1,R_2$.
$\mu_{k,i}\ (\mu^\vee_{k,j})$ denotes the length of $i$th row
($j$th column) of the Young diagram of $R_k$.
The sum is taken over all possible partitions $\bfR$
(including the empty partition).
The set of indices $(i,j)$ runs over the coordinates of
all boxes in the Young diagram of $R_k$.

The first two terms of $F^\pert_\YM$ represent
the contribution from the vector multiplet,
where the argument in the second term should be shifted
as above (see e.g.~\cite{Alday:2009aq}).
The last term of $F^\pert_\YM$ represents the contribution
from the four massless fundamental matters. Note that
we have taken account of the shift of the mass parameters
by $(\epsilon_1+\epsilon_2)/2$ which is needed
in order for the partition function to
have good modular properties \cite{Huang:2011qx}.
It is now well known 
that the UV gauge coupling $\tau_\UV$ appearing in
the Nekrasov partition function through $q_0=\exp(2\pi i\tau_\UV)$
is not identical with
the IR gauge coupling
$\tau$ represented by the complex modulus
of the Seiberg--Witten curve \cite{Dorey:1996bn}.
$q_0$ and $q=\exp(2\pi i\tau)$
are related to each other by \cite{Grimm:2007tm}
\eqb
q_0\Eqn{=}\frac{\varth_2^4}{\varth_3^4}\nn\\
\Eqn{=}
16q^{1/2}-128q+704q^{3/2}-3072q^2
+{\cal O}(q^{5/2}),
\eqe
or
\eqb
\label{qinq0}
q\Eqn{=}\frac{1}{256}q_0^2+\frac{1}{256}q_0^3+\frac{29}{8192}q_0^4
 +\frac{13}{4096}q_0^5+{\cal O}(q_0^6).
\eqe

The free energy of the above partition function can be expanded as
\eqb\label{FYM}
F_\YM\equiv\ln Z_\YM
\Eqn{=}\sum_{n,g=0}^\infty
  (\epsilon_1+\epsilon_2)^{2n}(-\epsilon_1\epsilon_2)^{g-1}
  (-a^2)^{-n-g+1}F_\YM^{(n,g)}.
\eqe
One can check that
\[\label{FYMlow}
F_\YM^{(0,0)}=\ln\frac{q_0}{q^{1/2}},\qquad
F_\YM^{(1,0)}=-\frac{1}{2}\ln 2a+\ln\frac{\varth_3^2}{\varth_4}
  +\frac{\ln 2}{3},\qquad
F_\YM^{(0,1)}=-\frac{1}{2}\ln 2a+\frac{2\ln 2}{3}.
\]
The second term $\ln\left(\varth_3^2/\varth_4\right)$ in
the expression of $F_\YM^{(1,0)}$ may be regarded
as the `$\grp{U}(1)$ factor,'
which should be stripped off
when the difference between
$\grp{SU}(2)$ and $\grp{U}(2)$ gauge groups matters
\cite{Alday:2009aq}.\footnote{
To be precise, the $\grp{U}(1)$ factor proposed in \cite{Alday:2009aq}
increases $F_\YM^{(1,0)}$ by
$\ln\left(\varth_3^2/\varth_4^2\right)$.
This is very similar, but not identical to
the term $\ln\left(\varth_3^2/\varth_4\right)$ in (\ref{FYMlow}).
(We verified (\ref{FYMlow}) up to order $q_0^{15}$.)}
However, such a difference within
$F_\YM^{(1,0)}$ is not relevant to our discussion below.
It is known that
$F_\YM^{(n,g)}\ (n+g\ge 2)$ is a quasi modular form in $\tau$
of weight $2n+2g-2$ \cite{Grimm:2007tm}.
Explicit forms of them with small $n+g$
are found as \cite{Huang:2011qx}
\begin{align}
\hspace{5em}
F_\YM^{(2,0)}&=\frac{E_2}{96},\qquad
F_\YM^{(1,1)}=\frac{E_2}{24},\qquad
F_\YM^{(0,2)}=\frac{E_2}{32},\hspace{-10em}&\nn\\
F_\YM^{(3,0)}&=\frac{5E_2^2+13E_4}{11520},&
F_\YM^{(2,1)}&=\frac{10E_2^2+17E_4}{2880},
\hspace{5em}\nn\\
F_\YM^{(1,2)}&=\frac{95E_2^2+94E_4}{11520},&
F_\YM^{(0,3)}&=\frac{2E_2^2+E_4}{384}.
\end{align}
%

%%%
\subsection{Limit}
%%%

We claim that the above Nekrasov partition function $Z_\YM$
is reproduced from the reduced BPS index $Z$
by rescaling the variables as
\[\label{replacement}
\phi\to\beta a,\qquad
\epsilon_\alpha\to -i\beta\epsilon_\alpha
\]
and taking the limit $\beta\to 0$.
Recall that the free energy $F=\ln Z$
admits the genus expansion (\ref{genusexp}) with 
coefficients in the form (\ref{Fngform}).
By taking the above limit,
the higher genus part of the free energy becomes
\eqb
F\big|_{n+g\ge 2}\Eqn{=}
\sum_{n+g\ge 2}
 (\epsilon_1+\epsilon_2)^{2n}(-\epsilon_1\epsilon_2)^{g-1}
(-i\beta)^{2n+2g-2}
\sum_{h=1}^{n+g-1}
 \frac{F^{(n,g,h)}(\tau)}{\sinh^{2h}\beta a}\nn\\[1ex]
\Eqn{\stackrel{\hspace{-1em}\beta\to 0\hspace{-1em}}{\to}}
\sum_{n+g\ge 2}
 (\epsilon_1+\epsilon_2)^{2n}(-\epsilon_1\epsilon_2)^{g-1}
(-a^2)^{-n-g+1}F^{(n,g,n+g-1)}.
\eqe
This takes the same form as (\ref{FYM}).
We find that
\[\label{Nf4conjecture}
\fbox{$\displaystyle
F^{(n,g,n+g-1)}(\tau)=F_\YM^{(n,g)}(\tau)$}
\]
for $n+g\ge 2$.
We verified this equality
for $2\le n+g\le 6$ by explicit calculation.
We do not have a complete proof of this identity.
In the rest of this section
we present some evidence supporting further
this conjectured identity.

Let us first show that
the modular anomaly equation (\ref{MAEforFng})
reduces to that for the $\Nf=4$ theory.
We substitute the general form of $F^{(n,g)}$ (\ref{Fngform})
and $\phi = \beta a$ into (\ref{MAEforFng}).
We then multiply $\beta^{2n+2g-2}$ to both sides
and take the limit $\beta\to 0$.
The equation becomes
\eqb
\frac{\partial}{\partial E_2}\frac{F^{(n,g,n+g-1)}}{a^{2n+2g-2}}
\Eqn{=}
 \frac{1}{24}\sum_{n_1=0}^n\sum_{g_1=0}^g
\left(
 \frac{\partial}{\partial a}
 \frac{F^{(n_1,g_1,n_1+g_1-1)}}{a^{2n_1+2g_1-2}}\right)
\left(
 \frac{\partial}{\partial a}
 \frac{F^{(n-n_1,g-g_1,n-n_1+g-g_1-1)}}
      {a^{2n-2n_1+2g-2g_1-2}}\right)\nn\\
&&
+\frac{1}{24}\frac{\partial^2}{\partial a^2}
 \frac{F^{(n,g-1,n+g-2)}}{a^{2n+2g-4}}.
\eqe
Here $F^{(n,g,n+g-1)}$ for $n+g\le 1$ are understood as
\[
 F^{(0,0,-1)}=0,\qquad
 F^{(1,0,0)}
=F^{(0,1,0)}
=-\frac{1}{2}\ln 2a-\frac{1}{2}\ln\beta.
\]
Following \cite{Huang:2011qx},
we introduce the rescaled amplitudes $f^{(n,g)}$ by
\eqb
f^{(0,0)}\Eqn{=}0,\qquad
f^{(1,0)}=f^{(0,1)}=\frac{1}{4},\nn\\[1ex]
f^{(n,g)}\Eqn{=}(n+g-1) F^{(n,g,n+g-1)}\qquad
  \mbox{for}\quad n+g\ge 2
\eqe
and $f^{(n,g)}=0$ if $n<0$ or $g<0$.
The above equation is then written in terms of $f^{(n,g)}$
as
\[
\partial_{E_2}f^{(n,g)}
=\frac{n+g-1}{6}\left[
 \sum_{n_1=0}^n\sum_{g_1=0}^g f^{(n_1,g_1)}f^{(n-n_1,g-g_1)}
 +\left(n+g-\frac{3}{2}\right)f^{(n,g-1)}\right].
\]
This is precisely the modular anomaly equation
for the $\Nf=4$ theory \cite{Huang:2011qx}.

Let us next show that the perturbative part of
the BPS index of E-strings
reproduces that of the $\Nf=4$ theory in the limit $\beta\to 0$.
By starting from the expression (\ref{Fpertingamma})
with (\ref{gamma5d}), it is not difficult to show that
\eqb
\lefteqn{
\lim_{\beta\to 0}
  F^\pert(\beta a,-i\beta\epsilon_1,-i\beta\epsilon_2)}\nn\\
\Eqn{=}
 -\gamma^\YM_{\epsilon_1,\epsilon_2}(2a)
 -\gamma^\YM_{\epsilon_1,\epsilon_2}(2a-\epsilon_1-\epsilon_2)
 +8\gamma^\YM_{2\epsilon_1,2\epsilon_2}(2a-\epsilon_1-\epsilon_2)
+\frac{\epsilon_1^2+\epsilon_1\epsilon_2+\epsilon_2^2}
      {2\epsilon_1\epsilon_2}\ln\beta.\nn\\
\eqe
Here $\ln \beta$ is understood as an infinite constant.
By performing the series expansion in $\epsilon_1,\epsilon_2$,
it is easy to check that
\[
\gamma^\YM_{2\epsilon_1,2\epsilon_2}(2a-\epsilon_1-\epsilon_2)
=\gamma^\YM_{\epsilon_1,\epsilon_2}
   \left(a-\frac{\epsilon_1}{2}-\frac{\epsilon_2}{2}\right)
-\frac{\ln 2}{2}\frac{a^2}{\epsilon_1\epsilon_2}
+\frac{\ln 2}{24}
 \frac{\epsilon_1^2+\epsilon_2^2}{\epsilon_1\epsilon_2}.
\]
Therefore,
$F^\pert$ in the $\beta\to 0$ limit reproduces $F^\pert_\YM$
\[
\lim_{\beta\to 0}
  F^\pert(\beta a,-i\beta\epsilon_1,-i\beta\epsilon_2)
=F^\pert_\YM+\mbox{\it shifts}.
\]
Here {\it shifts} denote
constant shifts of $F_\YM^{(n,g)}$ with $n+g\le 1$.
Hence, the conjectured identity (\ref{Nf4conjecture})
in the limit $q\to 0$ has been proved.

Finally, let us verify that
the instanton expansion performed in section~4
is also consistent with the Nekrasov partition function.
One can define the counterpart of $F^\qinst_k$
studied in section~4.
As we define $Z$ as the BPS index
rather than the topological string partition function,
$Z$ does not have the `classical part', as opposed to $Z_\YM$.
In order to make the comparison accurate,
we have to subtract such part at low genus from $F_\YM$.
More specifically, we define $F^\qinst_{\YM,k}$ by
\begin{align}
F^\inst_\YM\big|_{n+g\ge 2}
&=\ln Z^\inst_\YM
 -\frac{a^2}{\epsilon_1\epsilon_2}\ln\frac{q_0}{2^4q^{1/2}}
 +\frac{(\epsilon_1+\epsilon_2)^2}{\epsilon_1\epsilon_2}
  \ln\frac{\varth_3^2}{\varth_4}\nn\\
&=:\sum_{k=1}^\infty q^k F^\qinst_{\YM,k}(a,\epsilon_1,\epsilon_2).
\end{align}
By evaluating the Nekrasov partition function (\ref{ZYMinst})
up to and including the terms of order $q_0^5$,
we obtain
\eqb
F^\qinst_{\YM,1}(a,\hbar,-\hbar)
 \Eqn{=}\frac{12\hbar^2a^2}{(2a-\hbar)^2(2a+\hbar)^2},\\
F^\qinst_{\YM,2}(a,\hbar,-\hbar)
 \Eqn{=}
 \frac{18\hbar^2a^2
       (512a^8-1024\hbar^2a^6+1104\hbar^4a^4-584\hbar^6a^2+27\hbar^8)}
      {(2a-\hbar)^4(2a+\hbar)^4(2a-3\hbar)^2(2a+3\hbar)^2}.\qquad
\eqe
One can easily check that $F^\qinst_k$ presented
in (\ref{Fqinst1}), (\ref{Fqinst2}) indeed reduce to
the above expressions
\[
\lim_{\beta\to 0}F^\qinst_k(\beta a,-i\beta\hbar,i\beta\hbar)
=F^\qinst_{\YM,k}(a,\hbar,-\hbar).
\]
%

%%%%%%%%%%%%%%%%%%%%%%%%%%%%%%%%%%%%%%%%%%%%%%%%%%%%%%%%%%%%%%%%%%%%%%%%
\section{Discussion}
%%%%%%%%%%%%%%%%%%%%%%%%%%%%%%%%%%%%%%%%%%%%%%%%%%%%%%%%%%%%%%%%%%%%%%%%

In this paper we studied
a reduced BPS index of E-strings.
This is obtained by evaluating the refined BPS index of E-strings
at special values of $E_8$ Wilson line parameters
that correspond to breaking the global $E_8$ symmetry down to
$D_4\oplus D_4$.
The index admits three kinds of expansions,
each of which corresponds to an entirely different physical picture.
We clarified the structure of expansions and
determined some expansion coefficients.
We elucidated in detail how the reduced BPS index reduces to
the Nekrasov partition function
for 4d ${\cal N}=2$ supersymmetric $\grp{SU}(2)$
gauge theory with $\Nf=4$ massless flavors.

It would be interesting to generalize the present study
to the case of the $\grp{SU}(2)$ $\Nf=4$ theory
with massive flavors. 
At the level of the Seiberg--Witten curve,
it is already known \cite{Sakai:2012ik}
how to identify the $E_8$ Wilson line parameters
with the masses of flavors in the $\grp{SU}(2)\ \Nf=4$ theory.
Interestingly, the identification has to take a quite
nontrivial form in order to keep the modular properties intact.

A very simple form of 
Nekrasov-type expression is available for
the Seiberg--Witten prepotential for the E-string theory
\cite{Sakai:2012zq,Sakai:2012ik,Ishii:2013nba}.
The expression however
does not reproduce the higher genus part.
An appropriate modification or some alternative formula
is anticipated.
The present study may open up a way
of constructing
the all genus Nekrasov-type partition function
for the E-string theory.

The reduced BPS index of E-strings gives a novel
trigonometric generalization of the Nekrasov partition function
for the $\grp{SU}(2)\ \Nf=4$ theory.
As is well known,
this Nekrasov partition function is identified with
a certain Virasoro conformal block through
Alday--Gaiotto--Tachikawa relation \cite{Alday:2009aq}.
It would be extremely interesting if there exists
an interpretation of the BPS index of E-strings on the CFT side.

%%%%%%%%%%%%%%%%%%%%%%%%%%%%%%%%%%%%%%%%%%%%%%%%%%%%%%%%%%%%%%%%%%%%%%%%
\vspace{3ex}

\begin{center}
  {\bf Acknowledgments}
\end{center}

This work was supported in part by
JSPS KAKENHI Grant Number 26400257
and JSPS Japan--Hungary Research Cooperative Program.

\vspace{3ex}

%%%%%%%%%%%%%%%%%%%%%%%%%%%%%%%%%%%%%%%%%%%%%%%%%%%%%%%%%%%%%%%%%%%%%%%%
%%% Appendices %%%

\appendix

%%%%%%%%%%%%%%%%%%%%%%%%%%%%%%%%%%%%%%%%%%%%%%%%%%%%%%%%%%%%%%%%%%%%%%%%
\section{Conventions of special functions and useful formulas}
%%%%%%%%%%%%%%%%%%%%%%%%%%%%%%%%%%%%%%%%%%%%%%%%%%%%%%%%%%%%%%%%%%%%%%%%

The Jacobi theta functions are defined as
\eqb
\varth_1(z,\tau)\Eqn{:=}
   i\sum_{n\in \bbZ} (-1)^n y^{n-1/2}q^{(n-1/2)^2/2},\\
\varth_2(z,\tau)\Eqn{:=}
   \sum_{n\in \bbZ} y^{n-1/2}q^{(n-1/2)^2/2},\\
\varth_3(z,\tau)\Eqn{:=}
   \sum_{n\in \bbZ} y^n q^{n^2/2},\\
\varth_4(z,\tau)\Eqn{:=}
   \sum_{n\in \bbZ} (-1)^n y^n q^{n^2/2},
\eqe
where
\[
y=e^{i z},\qquad q=e^{2\pi i \tau}.
\]
In this paper we adopt slightly different convention
for the theta functions as
compared to our previous works
\cite{Sakai:2011xg,Sakai:2012zq,Sakai:2012ik,Ishii:2013nba}.
The above theta functions are related to
those in our previous convention by
\[
\varth_k(z,\tau)=
\varth^{\mbox{\scriptsize previous}}_k\left(\frac{z}{2\pi},\tau\right).
\]
We often use the following
abbreviated notation
\[
\varth_k(z) := \varth_k(z,\tau),\qquad
\varth_k := \varth_k(0,\tau).
\]

The Dedekind eta function is defined as
\[
\eta(\tau) := q^{1/24}\prod_{n=1}^\infty (1-q^n).
\]
The Eisenstein series are given by
\[
E_{2n}(\tau)
   =1-\frac{4n}{B_{2n}}
   \sum_{k=1}^{\infty}\frac{k^{2n-1}q^k}{1-q^k}
\]
for $n\in\bbZ_{>0}$.
The Bernoulli numbers $B_k$ are defined by
\eqb
\frac{x}{e^x-1}\Eqn{=}\sum_{k=0}^\infty\frac{B_k}{k!}x^k.
\eqe
We often abbreviate $\eta(\tau),\,E_{2n}(\tau)$ as $\eta,\,E_{2n}$
respectively.

The Weierstrass elliptic functions
are defined as
\eqb
\sigma(z;2\omega_1,2\omega_3)
  \Eqn{:=}z\hspace{-1em}
  \prod_{(m,n)\in\bbZ^2_{\ne (0,0)}}\hspace{-.5em}
  \left(1-\frac{z}{\Omega_{m,n}}\right)
  \exp\left[\frac{z}{\Omega_{m,n}}+\frac{z^2}{2{\Omega_{m,n}}^2}
      \right],\\
\zeta(z;2\omega_1,2\omega_3)
  \Eqn{:=}\frac{1}{z}
  +\sum_{(m,n)\in\bbZ^2_{\ne (0,0)}}
  \left[\frac{1}{z-\Omega_{m,n}}+\frac{1}{\Omega_{m,n}}
    +\frac{z}{{\Omega_{m,n}}^2}\right],\\
\wp(z;2\omega_1,2\omega_3)
  \Eqn{:=}\frac{1}{z^2}
  +\sum_{(m,n)\in\bbZ^2_{\ne (0,0)}}
  \left[\frac{1}{(z-\Omega_{m,n})^2}
    -\frac{1}{{\Omega_{m,n}}^2}\right],
\eqe
where $\Omega_{m,n}=2m\omega_1 + 2n\omega_3$.
In this paper we always set
$2\omega_1=2\pi,\ 2\omega_3=2\pi\tau$
and use the following abbreviated notation
\[
\sigma(z):=\sigma(z;2\pi,2\pi\tau),\qquad
\zeta(z):=\zeta(z;2\pi,2\pi\tau),\qquad
\wp(z):=\wp(z;2\pi,2\pi\tau).
\]
Note that
\[
\frac{d}{dz}\ln\sigma(z)=\zeta(z),\qquad
\frac{d}{dz}\zeta(z)=-\wp(z).
\]
The sigma function $\sigma(z)$
is related to the Jacobi theta function $\varth_1(z)$ by
\[\label{thetasigma}
\varth_1(z)=e^{-\frac{1}{24}E_2 z^2}\eta^3\sigma(z).
\]

To expand the above elliptic functions in $q=e^{2\pi i\tau}$,
the following formulas are useful:
\eqb
\label{theta_qexp}
\varth_1(z)
  \Eqn{=}2q^{1/12}\eta(\tau)\sin\frac{z}{2}
  \prod_{n=1}^\infty\left(1-2q^n\cos z+q^{2n}\right),\\
\label{zeta_qexp}
\zeta(z)
\Eqn{=}\frac{z}{12}
 +\frac{1}{2}\cot\frac{z}{2}
 +2\sum_{n=1}^\infty \frac{q^n}{1-q^n}\left(\sin nz - nz\right),\\
\label{wp_qexp}
\wp(z)
\Eqn{=}-\frac{1}{12}
 +\frac{1}{4\sin^2(z/2)}
 +4\sum_{n=1}^\infty \frac{n q^n}{1-q^n}\sin^2\frac{nz}{2}.
\eqe
The expansions of the elliptic functions in $z$ about $z=0$ 
are obtained as
\eqb
\label{sigma_zexp}
\sigma(z)\Eqn{=}z\exp\left(-\sum_{n=1}^\infty
  \frac{c_n}{(2n+1)(2n+2)}z^{2n+2}\right),\\
\label{zeta_zexp}
\zeta(z)\Eqn{=}\frac{1}{z}-\sum_{n=1}^\infty
  \frac{c_{n}}{2n+1}z^{2n+1},\\
\label{wp_zexp}
\wp(z)\Eqn{=}\frac{1}{z^2}+\sum_{n=1}^\infty c_{n}z^{2n},
\eqe
where $c_n$ are determined by the recurrence relation
\eqb
c_1 \Eqn{=} \frac{E_4}{240},\qquad
c_2 =       \frac{E_6}{6048},\nn\\
c_n \Eqn{=} \frac{3}{(n-2)(2n+3)}\sum_{k=1}^{n-2}c_k c_{n-k-1}
\qquad (n\ge 3).
\label{cnrecur}
\eqe
%

%%%%%%%%%%%%%%%%%%%%%%%%%%%%%%%%%%%%%%%%%%%%%%%%%%%%%%%%%%%%%%%%%%%%%%%%
%%%%%%%%%%%%%%%%%%%%%%%%%%%%%%%%%%%%%%%%%%%%%%%%%%%%%%%%%%%%%%%%%%%%%%%%
%%%%%%%%%%%%%%%%%%%%%%%%%%%%%%%%%%%%%%%%%%%%%%%%%%%%%%%%%%%%%%%%%%%%%%%%
%%% References %%%

\renewcommand{\section}{\subsection}
\renewcommand{\refname}

\end{document}